\documentclass[12pt, epsfig]{article}
\usepackage{graphicx}
  \usepackage{amsthm,amsfonts}
  \usepackage{amsmath}
  \usepackage{bm}
\newcommand{\bea}   {\begin{eqnarray}}
\newcommand{\eea}   {\end{eqnarray}}
\def\Pop#1{P^{(#1)}}
\def\Xop#1{X^{(#1)}}

\def\Jop#1{J^{(#1)}}

\begin{document}
\renewcommand{\thefootnote}{\fnsymbol{footnote}}

\thispagestyle{empty}

\title{Chiral and Real ${\cal N}=2$ supersymmetric \\${\ell}$-conformal Galilei algebras}

\author{N. Aizawa\thanks{{\em e-mail: aizawa@mi.s.osakafu-u.ac.jp}}, \quad
Z. Kuznetsova\thanks{{\em e-mail: zhanna.kuznetsova@ufabc.edu.br}}
\quad and\quad F.
Toppan\thanks{{\em e-mail: toppan@cbpf.br}}
\\
\\
}
\maketitle

\centerline{$^{\ast}${\it Dep. Math. and Inform. Sciences, Grad. School of Science,}}
{\centerline {\it Osaka Prefecture Univ., Nakamozu Campus, Sakai,}}
{\centerline {\it\quad
Osaka 599-8351, Japan.}
\centerline{$^{\dag}${\it UFABC, Rua Santa Ad\'elia 166, Bangu,}}{\centerline {\it\quad
cep 09210-170, Santo Andr\'e (SP), Brazil.}
\centerline{$^{\ddag}${\it CBPF, Rua Dr. Xavier Sigaud 150, Urca,}}{\centerline {\it\quad
cep 22290-180, Rio de Janeiro (RJ), Brazil.}
~\\
\maketitle
\begin{abstract}
~\\
Inequivalent ${\cal N}=2$ supersymmetrizations of the ${\ell}$-conformal Galilei algebra 
in $d$-spatial dimensions 
are constructed from the chiral $(2,2)$ and the real $(1,2,1)$ basic supermultiplets of the ${\cal N}=2$ supersymmetry. For non-negative integer and half-integer ${\ell}$ both superalgebras admit a consistent truncation with a (different) finite number of generators. 
The real ${\cal N}=2$ case coincides with the superalgebra
introduced by Masterov, while the chiral ${\cal N}=2$ case is a new superalgebra.

 We present $D$-module representations of both superalgebras. 
Then we investigate the new superalgebra derived from the chiral supermultiplet. 
It is shown that it admits two types of central extensions, one is found for any $d$ and half-integer ${\ell }$ and the other only for $d=2$ and integer $ \ell. $ 
For each central extension the centrally extended $\ell$-superconformal Galilei algebra is realized in terms of its super-Heisenberg subalgebra generators.

\end{abstract}
\vfill

\rightline{CBPF-NF-003/13}

\newpage
\section{Introduction}

In this paper we present the ${\cal N}=2$ supersymmetrizations of the ${\ell}$-conformal Galilei algebras in $d$ space dimensions. 
Here $ \ell $ is the parameter characterising the structure of the algebra and takes a non-negative integer or half-integer value. 
We construct at first the two finite, centerless, Lie superalgebras which are associated with the ${\cal N}=2$ chiral and real representations, respectively, and give their $D$-module representations. 
We prove that the centerless, finite
${\cal N}=2$ ${\ell}$-superconformal Galilei algebra of ref. \cite{mas}, expressed in terms of superfields, 
corresponds to the particular choice of the ${\cal N}=2$ real representation. The novelty here is the introduction
of the second superalgebra, associated with the ${\cal N}=2$ chiral representation. In \cite{aiz} the (mass and exotic) central extensions of the  superalgebra of ref. \cite{mas} were given.
We prove that mass and exotic central extensions also exist for the chiral supersymmetrization.  \par
Our construction of the $D$-module reps for the cases at hand is based on the classified $D$-module reps for the relevant subalgebras, such as the ${\cal N}$-extended supersymmetry in $0+1$ dimensions (see \cite{pt} and \cite{krt})
and the $D$-module reps of the finite, simple, ${\cal N}$-extended one-dimensional superconformal algebras (see \cite{kuto} and \cite{khto}). \par
For ${\cal N}=2$ the chiral and real representations of the $0+1$-dimensional supersymmetry are also denoted as ``$(2,2)$" and ``$(1,2,1)$" reps, as discussed in Section {\bf 2}. 
For this reason the chiral and real ${\cal N}=2$ supersymmetrizations of the ${\ell} $-conformal Galilei algebra are denoted as ${\cal G}_{(2,2)}$ and ${\cal G}_{(1,2,1)}$, respectively. 
For a non-negative integer or half-integer value of the parameter ${\ell}$ they admit a consistent truncation as non semi-simple, finite, centerless, Lie superalgebras.
The non-vanishing (anti)commutators of ${\cal G}_{(2,2)}$ are presented in formulas (\ref{sl21}) and (\ref{g22extra}). The non-vanishing (anti)commutators of ${\cal G}_{(1,2,1)}$ are given in (\ref{sl21}) and (\ref{g121extra}).
Their respective $D$-module representations (realized by differential operators in $d+1$ space-time dimensions) are given in ({\ref{g22dmod}) and (\ref{g121dmod}).\par
It follows from the Section {\bf 2} results that, for ${\ell}\geq \frac{1}{2}$, the ${\cal G}_{(2,2)}$ 
superalgebra contains a larger number (=$2d$) of even generators with respect to the ${\cal G}_{(1,2,1)}$ superalgebra.\par
Concerning the central extensions the results are common for ${\cal G}_{(2,2)}$ and ${\cal G}_{(1,2,1)}.$ 
The superalgebra ${\cal G}_{(2,2)}$ in $d$ dimensions has a central extension if $ \ell $ is a half-integer. 
This extension is called the {\em mass extension} since the eigenvalue of the central element, in the case of 
bosonic conformal Galilei algebra with $ \ell=1/2,$ can be interpreted 
as the mass of the free Schr\"odinger equation. 
${\cal G}_{(2,2)}$ in 2 dimensions has another type of central extension if $ \ell $ is an integer. 
In analogy with the ${\cal G}_{(1,2,1)}$ case it will be called the {\em exotic central extension}. Each central extension makes an (anti)commutative subalgebra of ${\cal G}_{(2,2)}$ 
noncommutative. This noncommutative subalgebra is the super-Heisenberg algebra. 
Representations of the centrally extended ${\cal G}_{(2,2)}$ algebras are given
in terms of the super-Heisenberg algebra generators.\par
The supersymmetrization of the $d$-dimensional conformal Galilei algebras has a long history. 
For ${\ell}=\frac{1}{2}$, the so called Schr\"odinger algebra, there are many works on supersymmetrizations.  
For instance, in \cite{dh}  Duval and Horv\'athy presented a systematic way to supersymmetrize the Schr\"odinger algebra. 
Sakaguchi and Yoshida identified various supersymmetrizations as subalgebra of superconformal algebras \cite{SY1,SY2} 
(see also \cite{aiz} for a more complete list of references).  
For other values of the parameter ${\ell}$, the investigation of the supersymmetric cases started recently. 
Various $ \ell =1 $ extensions were considered in \cite{AzLu,Sakaguchi,BaMa,FeLu,Mandal}. 
Only the $ {\cal N} = 2 $ extensions have been studied for arbitrary values of $ \ell $ \cite{mas,aiz}. 

Supersymmetrizations of the ${\ell}$-conformal Galilei algebra can be extended for ${\cal N}>2$  
(for $ \ell = 1 $ see \cite{AzLu,Sakaguchi,FeLu}). 
The approach based on the $D$-module reps which we used for ${\cal N}=2$ can be applied for larger values of ${\cal N}$. Some new features, which will be discussed in an Appendix, appear. The most relevant is the existence, for ${\cal N}=4$, of a critical scaling dimension and the fact that, even for ${\ell}$ integer or half-integer, the resulting ${\cal N}=4$ ${\ell}$-conformal Galilei algebras are infinite-dimensional (the
one based on the $(4,4)$ rep, however, closes as a non-linear $W$-algebra on a finite number of generators).\par
The scheme of the paper is the following. In Section {\bf 2} we introduce the finite centerless Lie superalgebras ${\cal G}_{(2,2)}$ and ${\cal G}_{(1,2,1)}$ together with their $D$-module representations. 
The central extensions of ${\cal G}_{(2,2)}$ and the super-Heisenberg representations of the centrally extended  ${\cal G}_{(2,2)}$ superalgebras 
are presented in Section {\bf 3}. In the Conclusions we will comment about the
possible dynamical applications of the present results. Some issues concerning the supersymmetrization of the ${\ell}$-conformal Galilei algebras for ${\cal N}\geq 4$ are discussed in the Appendix.

\section{The centerless chiral and real ${\cal N}=2$ superconformal Galilei algebras and their $D$-module reps.}

The conformal Galilei algebras are non semi-simple Lie algebras specified by two parameters, $d$ taking a positive integer value (the dimensionality of the non-relativistic space) and  ${\ell}$ being a non-negative integer or half-integer \cite{NdORM}. 
The parameter $d$ characterizes the
maximal semisimple subalgebra $sl(2, {\mathbb R})\oplus so(d)$, while ${\ell}$ specifies the
$(2{\ell}+1)d$ dimensional abelian ideal which carries a spin ${\ell }$ representation of the $sl(2,{\mathbb R})$
subalgebra. \par
The ${\ell}$-conformal Galilei algebras can be naturally realized (and even defined) in terms of their $D$-module
representations, given by differential operators with respect to the time ($t$) and the space coordinates
($x_i$, with $i=1,\ldots , d$).\par
Their supersymmetric extensions can be defined by enlarging the set of the $D$-module operators by introducing the (operators associated to the) odd
generators, together with the extra generators which are required for the closure of the superalgebra as a graded Lie algebra.\par
The supersymmetrization of the ${\ell}$-conformal Galilei algebra must satisfy the constraints induced by the supersymmetrization of its bosonic subalgebras. In this respect two main features are crucial for implementing the whole procedure. The first one requires the knowledge of the $D$-module reps, given in \cite{{pt},{krt}}, of the algebra  of the ${\cal N}$-Extended Supersymmetric Mechanics in $(0+1)$-dimension 
(the algebra of the Supersymmetric Quantum Mechanics, see \cite{wit}).
The second feature is based on the construction of the $D$-module reps for the supersymmetric extensions of the
$sl(2,{\mathbb R})$ algebra (see \cite{kuto} and \cite{khto}), leading to the $D$-module reps of the finite,
simple, one-dimensional Superconformal Lie algebras.\par
For ${\cal N}=2$, there are two fundamental $D$-module reps of the algebra of the Supersymmetric Quantum Mechanics. They are denoted as $(2,2)$ and $(1,2,1)$, respectively. 
The meaning of this notation is the following. The operators of both representations act on $2$ bosonic and $2$ fermionic time-dependent fields, whose scaling dimensions, however, differ. In the first case both bosons have scaling dimension $\lambda$ and both fermions have scaling dimension $\lambda+\frac{1}{2}$.
In the second case the scaling dimension assignment is the following: $\lambda$ for the first boson, $\lambda+\frac{1}{2}$ for the two fermions and $\lambda+1$ for the remaining boson (known as the auxiliary field). In the superspace language the $(2,2)$ rep is known as the chiral representation, while the
$(1,2,1)$ rep is known as the real representation, see \cite{pt}.\par
The compatibility of these representations with the $D$-module rep of the $sl(2)$ algebra (for the correct assignment of the scaling dimension) induces, in both cases, a $D$-module rep for the ${\cal N}=2$ extension of the $sl(2)$ algebra, namely the $sl(2|1)$ superalgebra.  For any $\lambda$ we end up with either the $(2,2)$ chiral or the $(1,2,1)$ real  $D$-module rep of $sl(2|1)$ (see \cite{kuto} for details). \par
In order to get the supersymmetric extension of the conformal Galilei algebra we need to accommodate the generators $P^{(n)}_i$ (the basis of the abelian ideal of the bosonic conformal Galilei algebra) and $M_{ij}$ (spanning the $so(d)$
subalgebra). Their $D$-module representation, in the bosonic case, is known (see, e.g. \cite{{mas},{NdORM}}):
$P^{(n)}_i= t^n\partial_{x_i}$, $M_{ij}=x_i\partial_{x_j}-x_j\partial_{x_i}$. It is natural to assume that the generators $P^{(n)}_i$  and $M_{ij}$ act with the same transformations on each component field
of the given supermultiplet (in the Galilei case each component field $\phi$ carries a dependence on the time and space coordinates, so that $\phi\equiv \phi (t, x_i)$).\par
In the bosonic $D$-module rep the compatibility of the $sl(2)$ generators with the $P^{(n)}_i$, $M_{ij}$ generators requires that the dilatation operator $D$ and the operator $K$ (the conformal partner of the Hamiltonian $H$) are differential operators which,
unlike the $0+1$ conformal case, depend not just on $t$, but also on the space coordinate $x_i$'s. 
In the $0+1$-dimensional case, acting on a field with scaling dimension $\lambda$,  we have (see \cite{kuto})
\bea
& H=\partial_t, \quad D= -t\partial_t -\lambda, \quad K= -t^2\partial_t-2\lambda t.& \nonumber
\eea
In the $(d+1)$ dimensional case, in order to have consistent commutators with $P^{(n)}_i$ and $M_{ij}$, in the above formulas the scaling parameter $\lambda$ has to be replaced with the operator
${\widehat\lambda}$:
\bea\label{lambdahat}
\lambda &\mapsto& {\widehat\lambda} = \lambda +{\ell }\sum_{i=1}^d x_i\partial_{x_i}. 
\eea
In this way the second parameter ${\ell}$ (the one related to the spin) is introduced.\par
This construction, for the chiral and the real cases, produces the differential operators in $d+1$ dimensions
of the superalgebra $sl(2|1)$ (the generators $H,D,K$, the $R$-symmetry operator $R$, the supercharges $Q_a$
and their superconformal partners $S_a$, for $a=0,1$), together with the generators $P^{(n)}_i$, $M_{ij}$. The closure of the superconformal Galilei as a graded Lie algebra requires the introduction of extra generators arising from the (anti)commutators of the previous operators. We get, e.g., the fermionic operators $X^{(n)}_{a,i}$ from the commutators $[S_a, P^{(n)}_i]$ and the bosonic operators
$J_i^{(n)}$ from the set of anticommutators $\{S_a, X^{(n)}_{b,i}\}$. No further generator is required to close the superalgebra. Since the graded Jacobi identities are satisfied by construction, the supersymmetric extension, induced by the $D$-module, of the conformal Galilei algebra is a graded Lie algebra.\par
An unexpected feature emerges at this point. Two inequivalent supersymmetric extensions are obtained,
starting either from the chiral or from the real representation of the (one-dimensional) ${\cal N}=2$ supersymmetry.  The
associated superalgebras will be denoted as ${\cal G}_{(2,2)}$ and ${\cal G}_{(1,2,1)}$, respectively. The latter coincides with the centerless ${\cal N}=2$ superconformal Galilei algebra, expressed in terms of superfields, introduced by Masterov in \cite{mas}. The novelty here is that the Masterov's supersymmetrization is not unique, as the existence of the new supersymmetrization, ${\cal G}_{(2,2)}$, proves.\par
Both supersymmetrizations can be consistently truncated, for a non-negative integer or half-integer ${\ell}$,
to non semi-simple, finite-dimensional graded Lie algebras. The inequivalence of the two supersymmetrizations is recovered from their structure constants and the different total number of generators entering
${\cal G}_{(2,2)}$ and ${\cal G}_{(1,2,1)}$, respectively. Besides the $4$ bosonic and $4$ fermionic generators
entering the $sl(2|1)$ subalgebra and the  $\frac{d(d-1)}{2}$ $so(d)$ generators $M_{ij}$ we have, for both the truncated superalgebras ${\cal G}_{(2,2)}$ and ${\cal G}_{(1,2,1)}$,\\
the $P^{(n)}_i$ generators for $n=0,1, \ldots, 2{\ell }$ (their number is $d\times(2{\ell}+1)$) and \\
the $X_{a,i}^{(n)}$ generators for $n=0,1,\ldots, 2{\ell}-1$ and $a=0,1$ (their number is $2d\times 2{\ell}$).\\ For ${\cal G}_{(2,2)}$ the
$J_i^{(n)}$ generators are obtained for $n=0,1,\ldots, 2{\ell}$ (their number is $d\times (2{\ell}+1)$).\\
For ${\cal G}_{(1,2,1)}$ they are obtained for $n=0,1,\ldots, 2{\ell}-2$ (their number is $d\times (2{\ell}-1)$ for ${\ell }\geq \frac{1}{2}$).
\par
Therefore the ${\cal G}_{(2,2)}$ superalgebra contains, for ${\ell}\geq \frac{1}{2}$, $2d$ extra generators in the $J^{(n)}_i$ set
with respect to the ${\cal G}_{(1,2,1)}$ superalgebra.  Both ${\cal G}_{(2,2)}$ and ${\cal G}_{(1,2,1)}$ are centerless superalgebras which can accommodate central extensions, as shown in Section {\bf 3}.\par
We present here our results giving the structure constants and the $D$-module reps of the ${\cal G}_{(2,2)}$ and ${\cal G}_{(1,2,1)}$ superalgebras. Their common $sl(2|1)$ subalgebra reads as follows
\begin{eqnarray}\label{sl21}
\relax [D,H]&=& H,\nonumber\\
\relax [D,K]&=& -K,\nonumber\\
\relax [H,K]&=& 2D,\nonumber\\
\relax [D,Q_a]&=& \frac{1}{2}Q_a,\nonumber\\
\relax [K,Q_a]&=& S_a,\nonumber\\
\relax [H,S_a]&=& Q_a,\nonumber\\
\relax [D,S_a]&=& -\frac{1}{2}S_a,\nonumber\\
\relax \{Q_a,Q_b\}&=& 2\delta_{ab}H,\nonumber\\
\relax \{S_a,S_b\}&=& -2\delta_{ab}K,\nonumber\\
\relax \{Q_a,S_b\}&=&-2\delta_{ab}D+ \epsilon_{ab}R,\nonumber\\
\relax [R, Q_a]&=&-\epsilon_{ab} Q_b,\nonumber\\
\relax [R, S_a]&=&-\epsilon_{ab} S_b,
\end{eqnarray}
($a,b=0,1$ and $\epsilon_{01}=-\epsilon_{10}=1$),
while the other $sl(2|1)$  (anti)commutators are vanishing.
\par
For ${\cal G}_{(2,2)}$ we further have the remaining non-vanishing (anti)commutators 
\begin{eqnarray}\label{g22extra}
\relax [H,P^{(n)}_i]&=& nP^{(n-1)}_i,\nonumber\\
\relax [D,P^{(n)}_i]&=& -(n-{\ell})P^{(n)}_i,\nonumber\\
\relax [K, P^{(n)}_i]&=& -(n-2{\ell})P^{(n+1)}_i,\nonumber\\
\relax [Q_a, P^{(n)}_i] &=& n X_{a,i}^{(n-1)},\nonumber\\
\relax [S_a, P^{(n)}_i]&=& (n-2{\ell}) X_{a,i}^{(n)},\nonumber\\
\relax [R,P_i^{(n)}]&=& 2{\ell}J_i^{(n)},\nonumber\\
\relax [H,X_{a,i}^{(n)}]&=& nX_{a,i}^{(n-1)},\nonumber\\
\relax [D,X_{a,i}^{(n)}]&=& -(n-{\ell}+\frac{1}{2})X_{a,i}^{(n)},\nonumber\\
\relax [K,X_{a,i}^{(n)}]&=& -(n-2{\ell}+1)X_{a,i}^{(n+1)},\nonumber\\
\relax \{Q_a,X_{b,i}^{(n)}\}&=& \delta_{ab}P_i^{(n)} -\epsilon_{ab}J_i^{(n)},\nonumber\\
\relax \{S_a,X_{b,i}^{(n)}\}&=& \delta_{ab}P_i^{(n+1)}-\epsilon_{ab}J_i^{(n+1)},\nonumber\\
\relax [R, X_{a,i}^{(n)}] &=& -(2{\ell}+1)\epsilon_{ab}X_{b,i}^{(n)},\nonumber\\
\relax [H,J_i^{(n)}]&=& nJ^{(n-1)}_i,\nonumber\\
\relax [D,J_i^{(n)}]&=& -(n-{\ell})J^{(n)}_i,\nonumber\\
\relax [K, J^{(n)}_i]&=& -(n-2{\ell})J^{(n+1)}_i,\nonumber\\
\relax [Q_a, J^{(n)}_i] &=& -\epsilon_{ab} n X_{b,i}^{(n-1)},\nonumber\\
\relax [S_a, J^{(n)}_i]&=& -\epsilon_{ab}(n-2{\ell}) X_{b,i}^{(n)},\nonumber\\
\relax [R,J_i^{(n)}]&=& -2{\ell}P^{(n)}_i,\nonumber\\
\relax [M_{ij},M_{kl}]&=& \delta_{jk}M_{il}+\delta_{il}M_{jk}+\delta_{jl}M_{ki}+\delta_{ik}M_{lj},\nonumber\\
\relax[P^{(n)}_k, M_{ij}] &=& \delta_{ik}P_j^{(n)}-\delta_{jk}P_i^{(n)},\nonumber\\
\relax [X_{a,k}^{(n)},M_{ij}]&=& \delta_{ik}X_{a,j}^{(n)}-\delta_{jk}X_{a,i}^{(n)},\nonumber\\
\relax [ J_k^{(n)}, M_{ij}]&=& \delta_{ik} J_j^{(n)}-\delta_{jk} J_i^{(n)}.
\eea
For ${\cal G}_{(1,2,1)}$ the further non-vanishing (anti)commutators are
\bea\label{g121extra}
\relax [H,P^{(n)}_i]&=& nP^{(n-1)}_i,\nonumber\\
\relax [D,P^{(n)}_i]&=& -(n-{\ell})P^{(n)}_i,\nonumber\\
\relax [K, P^{(n)}_i]&=& -(n-2{\ell})P^{(n+1)}_i,\nonumber\\
\relax [Q_a, P^{(n)}_i] &=& n X_{a,i}^{(n-1)},\nonumber\\
\relax [S_a, P^{(n)}_i]&=& (n-2{\ell}) X_{a,i}^{(n)},\nonumber\\
\relax [H,J^{(n)}_i]&=& nJ^{(n-1)}_i,\nonumber\\
\relax [D,J^{(n)}_i]&=& -(n-{\ell}+1)J^{(n)}_i,\nonumber\\
\relax [K, J^{(n)}_i]&=& -(n-2{\ell}+2)J^{(n+1)}_i,\nonumber\\
\relax [Q_a, J^{(n)}_i]&=& \epsilon_{ab}X_{b,i}^{(n)},\nonumber\\
\relax[S_a, J^{(n)}_i]&=& \epsilon_{ab} X_{b,i}^{(n+1)},\nonumber\\
\relax [H,X_{a,i}^{(n)}]&=& nX_{a,i}^{(n-1)},\nonumber\\
\relax [D,X_{a,i}^{(n)}]&=& -(n-{\ell}+\frac{1}{2})X_{a,i}^{(n)},\nonumber\\
\relax [K,X_{a,i}^{(n)}]&=& -(n-2{\ell}+1)X_{a,i}^{(n+1)},\nonumber\\
\relax [ R, X_{a,i}^{(n)}]&=& -\epsilon_{ab}X_{b,i}^{(n)},\nonumber\\
\relax \{Q_a,X_{b,i}^{(n)}\}&=& \delta_{ab}P^{(n)}_i +n\epsilon_{ab}J^{(n-1)}_i,\nonumber\\
\relax \{S_a,X_{b,i}^{(n)}\}&=& \delta_{ab}P^{(n+1)}_i+(n-2{\ell}+1)\epsilon_{ab}J^{(n)}_i,\nonumber\\
\relax [M_{ij},M_{kl}]&=& \delta_{jk}M_{il}+\delta_{il}M_{jk}+\delta_{jl}M_{ki}+\delta_{ik}M_{lj},\nonumber\\
\relax[P^{(n)}_k, M_{ij}] &=& \delta_{ik}P_j^{(n)}-\delta_{jk}P_i^{(n)},\nonumber\\
\relax [X_{a,k}^{(n)},M_{ij}]&=& \delta_{ik}X_{a,j}^{(n)}-\delta_{jk}X_{a,i}^{(n)},\nonumber\\
\relax [ J_k^{(n)}, M_{ij}]&=& \delta_{ik} J_j^{(n)}-\delta_{jk} J_i^{(n)}.
\eea
One should note the different structure constants (entering, for instance, the $\{Q_a, X^{(n)}_{b,i}\}$ anticommutator) with respect to the previous superalgebra.\newpage
Let $e_{IJ}$ denotes the supermatrix with entry $1$ at the crossing of the $I$-th row and $J$-th column and $0$ otherwise.\par
The ${\cal G}_{(2,2)}$ $D$-module rep is explicitly given by
\bea\label{g22dmod}
H&=& (e_{11}+e_{22}+e_{33}+e_{44})\partial_t,\nonumber\\
D&=& -(e_{11}+e_{22}+e_{33}+e_{44})(t\partial_t+\lambda +{\ell}x_i\partial_{x_i})-\frac{1}{2}(e_{33}+e_{44}),\nonumber\\
K&=& -(e_{11}+e_{22}+e_{33}+e_{44})(t^2\partial_t+2\lambda t+2{\ell}tx_i\partial_{x_i}) -t(e_{33}+e_{44}),\nonumber\\
R&=& -2({\ell}x_i\partial_{x_i}+\lambda)(e_{12}-e_{21}+e_{34}-e_{43})+(e_{34}-e_{43}),\nonumber\\
Q_0&=& e_{13}+e_{24}+(e_{31}+e_{42})\partial_t,\nonumber\\
Q_1&=& e_{14}-e_{23}-(e_{32}-e_{41})\partial_t,\nonumber\\
S_0&=& (e_{13}+e_{24})t+(e_{31}+e_{42})(t\partial_t+2\lambda+2{\ell} x_i\partial_{x_i}),\nonumber\\
S_1&=& (e_{14}-e_{23})t-(e_{32}-e_{41})(t\partial_t+2\lambda+2{\ell}x_i\partial_{x_i}),\nonumber\\
P^{(n)}_i&=& (e_{11}+e_{22}+e_{33}+e_{44})t^n\partial_{x_i},\nonumber\\
J^{(n)}_i&=& (e_{12}-e_{21}+e_{34}-e_{43})t^n\partial_{x_i},\nonumber\\
X_{0,i}^{(n)}&=& (e_{31}+e_{42})t^n\partial_{x_i},\nonumber\\
X_{1,i}^{(n)}&=& -(e_{32}-e_{41})t^n\partial_{x_i},\nonumber\\
M_{ij}&=& (e_{11}+e_{22}+e_{33}+e_{44})(x_i\partial_{x_j}-x_j\partial_{x_i}).
\eea 
The ${\cal G}_{(1,2,1)}$ $D$-module rep is explicitly given by
\bea\label{g121dmod}
H&=& (e_{11}+e_{22}+e_{33}+e_{44})\partial_t,\nonumber\\
D&=& -(e_{11}+e_{22}+e_{33}+e_{44})(t\partial_t+\lambda +{\ell}x_i\partial_{x_i})-\frac{1}{2}(2e_{22}+e_{33}+e_{44}),\nonumber\\
K&=& -(e_{11}+e_{22}+e_{33}+e_{44})(t^2\partial_t+2\lambda t+2{\ell} tx_i\partial_{x_i}) -t(2e_{22}+e_{33}+e_{44}),\nonumber\\
R&=& e_{34}-e_{43},\nonumber\\
Q_0&=& e_{13}+e_{42}+(e_{24}+e_{31})\partial_t,\nonumber\\
Q_1&=& e_{14}-e_{32}-(e_{23}-e_{41})\partial_t,\nonumber\\
S_0&=& (e_{13}+e_{42})t+(e_{24}+e_{31})(t\partial_t+2\lambda+2{\ell}x_i\partial_{x_i})+e_{24},\nonumber\\
S_1&=& (e_{14}-e_{23})t-(e_{23}-e_{41})(t\partial_t+2\lambda+2{\ell}x_i\partial_{x_i})-e_{23},\nonumber\\
P^{(n)}_i&=& (e_{11}+e_{22}+e_{33}+e_{44})t^n\partial_{x_i},\nonumber\\
J^{(n)}_i&=& e_{21}\cdot t^{n}\partial_{x_i},\nonumber\\
X_{0,i}^{(n)}&=& (e_{24}+e_{31})t^n\partial_{x_i},\nonumber\\
X_{1,i}^{(n)}&=& (e_{41}-e_{23})t^n\partial_{x_i},\nonumber\\
M_{ij}&=& (e_{11}+e_{22}+e_{33}+e_{44})(x_i\partial_{x_j}-x_j\partial_{x_i}).
\eea

\section{The centrally extended ${\cal G}_{(2,2)} $ superalgebras and their representations.}
 
 As shown in \cite{aiz}, $ {\cal G}_{(1,2,1)} $ has two types of central extensions according to 
the pair of values $(d,\ell). $ 
The central extensions make the (anti)commutative subalgebra spanned by $ \{ \Pop{n}_i, \Xop{n}_{a,i}, \Jop{n}_i \} $  
noncommutative. In \cite{aiz} the whole centrally extended $ {\cal G}_{(1,2,1)} $ algebra  is realized 
in terms of its super-Heisenberg subalgebra generators. 
In this section we show that the novel superalgebra $ {\cal G}_{(2,2)} $ shares the same properties. 
Computing the $ {\cal G}_{(2,2)} $ central extensions is an easy task. It is straightforward to verify that 
the following two types of central extensions are consistent with the graded Jacobi identities:
\begin{enumerate}
\renewcommand{\labelenumi}{(\roman{enumi})}
  \item for any $d$ and half-integer $\ell$ (mass central extension)
   \begin{eqnarray}
     & & [P_i^{(m)}, P_j^{(n)}] = \delta_{ij} \, \delta_{m+n,2\ell} I_m M, 
     \qquad 
      [ J_i^{(m)}, J_j^{(n)} ] = \delta_{ij} \,  \delta_{m+n,2\ell} I_m M, \label{Mext} \\
     & & \{ X_{a,i}^{(m)}, X_{b,j}^{(n)}\} = \delta_{ab} \,  \delta_{ij} \delta_{m+n,2\ell-1} \alpha_m M. \nonumber 
   \end{eqnarray}
  \item for $ d=2 $ and integer $ \ell $ (exotic central extension)
   \begin{eqnarray}
     & & [P_i^{(m)}, P_j^{(n)}] = \epsilon_{ij} \, \delta_{m+n,2\ell} I_m \Theta, 
         \qquad 
          [ J_i^{(m)}, J_j^{(n)} ] = \epsilon_{ij} \,  \delta_{m+n,2\ell} I_m \Theta, \nonumber \\
     & & \{ X_{a,i}^{(m)}, X_{b,j}^{(n)}\} = \delta_{ab} \,  \epsilon_{ij} \delta_{m+n,2\ell-1} \alpha_m \Theta,
     \quad (\epsilon_{12} = -\epsilon_{21} = 1). 
      \label{Thetaext}
   \end{eqnarray}
\end{enumerate}
The structure constants $ I_m, \; \alpha_m $ are common for the two extensions; they are given by
\begin{equation}
  I_m = c_0 (-1)^m m! (2\ell-m) !, \qquad \alpha_m = I_m/(2\ell-m), \label{one}
\end{equation}
where $ c_0 $ is an arbitrary number depending on $\ell$ but independent of $m.$

 We now turn to a representation of $ {\cal G}_{(2,2)} $ with the central extensions. 
Our aim is to express the generators of $ sl(2|1) \oplus so(d) $ in terms of $ \Pop{n}_i, \Xop{n}_{a,i}, \Jop{n}_i $ 
and $ M $ (or $\Theta$). 
This will be done by replacing the central elements $ M, \Theta $ with their eigenvalues $\mu, \theta.$ 
For simplicity we use the vector notations
\[
 \bm{P}^{(m)} = (\Pop{m}_1, \Pop{m}_2, \cdots, \Pop{m}_d),
 \qquad
 \bm{X}_a^{(m)} = (\Xop{m}_{a,1}, \Xop{m}_{a,2}, \cdots, \Xop{m}_{a,d}),
\] 
and introduce two types of product
\[
  \bm{P}^{(m)} \bm{J}^{(n)} = \sum_{i=1}^d \Pop{m}_i \Jop{n}_i, 
  \qquad
  \bm{P}^{(m)} \times \bm{J}^{(n)} = \sum_{i,j=1}^2 \epsilon_{ij} \Pop{m}_i \Jop{n}_j. 
\]
In these notations $ {\cal G}_{(2,2)} $ with the mass central extension is represented as follows:
\bea
  & & D = \frac{1}{2\mu}
      \left(
        \sum_{A=P,J} \sum_{m=0}^{2\ell} \frac{m-\ell}{I_m} \bm{A}^{(2\ell-m)} \bm{A}^{(m)}
        + \sum_{a=1,2} \sum_{m=0}^{2\ell-1} \frac{ m + \frac{1}{2}-\ell }{ \alpha_m } \bm{X}_a^{(2\ell-1-m)} \bm{X}_a^{(m)}
      \right),
   \nonumber \\
  & & H = -\frac{1}{2\mu}
      \left(
        \sum_{A=P,J} \sum_{m=0}^{2\ell} \frac{m}{I_m} \bm{A}^{(2\ell-m)} \bm{A}^{(m-1)}
        + \sum_{a=1,2} \sum_{m=0}^{2\ell-1} \frac{ m  }{ \alpha_m } \bm{X}_a^{(2\ell-1-m)} \bm{X}_a^{(m-1)}
      \right),
   \nonumber \\
  & & K = \frac{1}{2\mu}
      \left(
        \sum_{A=P,J} \sum_{m=0}^{2\ell} \frac{m}{I_m} \bm{A}^{(2\ell+1-m)} \bm{A}^{(m)}
        + \sum_{a=1,2} \sum_{m=0}^{2\ell-1} \frac{ m  }{ \alpha_m } \bm{X}_a^{(2\ell-m)} \bm{X}_a^{(m)}
      \right),
   \nonumber \\
  & & M_{ij} = \frac{1}{2\mu}
      \left(
        \sum_{A=P,J} \sum_{m=0}^{2\ell} \frac{1}{I_m} \Big( A_i^{(2\ell-m)} A_j^{(m)} - A_j^{(2\ell-m)} A_i^{(m)} \Big)
      \right.
   \nonumber \\
  & & \qquad \qquad \qquad \left.
      + \sum_{a=1,2} \sum_{m=0}^{2\ell-1} \frac{1}{ \alpha_m } 
        \Big( X_{a,i}^{(2\ell-1-m)} X_{a,j}^{(m)} + X_{a,i}^{(m)} X_{a,j}^{(2\ell-1-m)} \Big)
     \right),
   \nonumber \\
  & & R = \frac{1}{\mu} \left( 
        \sum_{m=0}^{2\ell} \frac{m}{I_m} \Big( \bm{P}^{(m)} \bm{J}^{(2\ell-m)} - \bm{P}^{(2\ell-m)} \bm{J}^{(m)} \Big)
      \right.
   \nonumber \\
  & & \qquad \qquad \qquad \left.
       + \sum_{m=0}^{2\ell-1} \frac{m+1}{\alpha_m} \Big( \bm{X}_1^{(m)} \bm{X}_2^{(2\ell-1-m)} + \bm{X}_1^{(2\ell-1-m)} \bm{X}_2^{(m)}  \Big)
      \right),
   \nonumber \\
  & & Q_a = -\frac{1}{\mu} 
        \sum_{m=0}^{2\ell} \frac{m}{I_m} \Big( \bm{P}^{(2\ell-m)} \bm{X}_a^{(m-1)} - \epsilon_{ab} \bm{J}^{(2\ell-m)} \bm{X}_b^{(m-1)} \Big),
   \nonumber \\
  & & S_a = -\frac{1}{\mu} 
        \sum_{m=0}^{2\ell} \frac{m}{I_m} \Big( \bm{P}^{(2\ell+1-m)} \bm{X}_a^{(m-1)} - \epsilon_{ab} \bm{J}^{(2\ell+1-m)} \bm{X}_b^{(m-1)} \Big).
   \label{MassRealization}
\eea
On the other hand $ {\cal G}_{(2,2)} $ with the exotic central extension is represented by
\bea
  & & D = \frac{1}{2\theta}
      \left(
        \sum_{A=P,J} \sum_{m=0}^{2\ell} \frac{\ell-m}{I_m} \bm{A}^{(2\ell-m)} \times \bm{A}^{(m)}
        + \sum_{a=1,2} \sum_{m=0}^{2\ell-1} \frac{ \ell- \frac{1}{2}-m }{ \alpha_m } \bm{X}_a^{(2\ell-1-m)} \times \bm{X}_a^{(m)}
      \right),
   \nonumber \\
  & & H = \frac{1}{2\theta}
      \left(
        \sum_{A=P,J} \sum_{m=0}^{2\ell} \frac{m}{I_m} \bm{A}^{(2\ell-m)} \times \bm{A}^{(m-1)}
        + \sum_{a=1,2} \sum_{m=0}^{2\ell-1} \frac{ m  }{ \alpha_m } \bm{X}_a^{(2\ell-1-m)} \times \bm{X}_a^{(m-1)}
      \right),
   \nonumber \\
  & & K = -\frac{1}{2\theta}
      \left(
        \sum_{A=P,J} \sum_{m=0}^{2\ell} \frac{m}{I_m} \bm{A}^{(2\ell+1-m)} \times \bm{A}^{(m)}
        + \sum_{a=1,2} \sum_{m=0}^{2\ell-1} \frac{ m  }{ \alpha_m } \bm{X}_a^{(2\ell-m)} \times \bm{X}_a^{(m)}
      \right),
   \nonumber \\
  & & M_{12} = \frac{1}{2\theta}
      \left(
        \sum_{A=P,J} \sum_{m=0}^{2\ell} \frac{1}{I_m} \bm{A}^{(2\ell-m)}  \bm{A}^{(m)}
        + \sum_{a=1,2} \sum_{m=0}^{2\ell-1} \frac{1}{ \alpha_m } \bm{X}_a^{(2\ell-1-m)} \bm{X}_a^{(m)}
      \right),
   \nonumber \\  
  & & R = \frac{1}{\theta} \left(
     \sum_{m=0}^{2\ell} \frac{m}{I_m} \Big( \bm{P}^{(2\ell-m)} \times \bm{J}^{(m)} + \bm{P}^{(m)} \times \bm{J}^{(2\ell-m)} \Big)
    \right. 
   \nonumber \\
  & & \qquad \qquad \qquad \left.
    + \sum_{m=0}^{2\ell-1} \frac{ m+1}{ \alpha_m } \Big( \bm{X}_1^{(m)} \times \bm{X}_2^{(2\ell-1-m)}
             - \bm{X}_1^{(2\ell-1-m)} \times \bm{X}_2^{(m)} \Big)
    \right),
  \nonumber \\
 & & Q_a = \frac{1}{\theta} 
        \sum_{m=0}^{2\ell} \frac{m}{I_m} \Big( \bm{P}^{(2\ell-m)} \times \bm{X}_a^{(m-1)} 
        - \epsilon_{ab} \bm{J}^{(2\ell-m)} \times \bm{X}_b^{(m-1)} \Big),
  \nonumber \\
 & & S_a = \frac{1}{\theta} 
        \sum_{m=0}^{2\ell} \frac{m}{I_m} \Big( \bm{P}^{(2\ell+1-m)} \times \bm{X}_a^{(m-1)} 
        - \epsilon_{ab} \bm{J}^{(2\ell+1-m)} \times \bm{X}_b^{(m-1)} \Big).
  \label{ExoticRealization}
\eea
One may verify by direct computation that the realizations (\ref{MassRealization}) and (\ref{ExoticRealization}) indeed satisfy the 
defining relations of their respective superalgebras.

\section{Conclusions.}

In this paper we proved the existence of two inequivalent ${\cal N}=2$ supersymmetrizations, the superalgebras ${\cal G}_{(2,2)}$ and ${\cal G}_{(1,2,1)}$, of the $d$-dimensional
${\ell}$-conformal Galilei algebras. They are recovered from, respectively, the chiral and the real representations of the ${\cal N}=2$ supersymmetry. Both superalgebras, for non-negative integer or half-integer values of ${\ell}$ are non semi-simple, finite, Lie superalgebras. Two types of central extensions, mass and exotic,
are encountered in both cases.\par
The present results and construction can be enlarged, in an essentially straightforward way, to the supersymmetrizations of the ${\ell}$-conformal Galilei algebras for ${\cal N}>2$. The basic tool is the use of
the fundamental $D$-module
representations of the ${\cal N}$-extended supersymmetry. New features, however, appear. The most relevant one (see the ${\cal N}=4$ example discussed in the Appendix) is that, even for positive integer or half-integer ${\ell}$, the resulting superconformal algebra closes, as a Lie superalgebra, on an infinite number of generators. It admits, nevertheless, a non-linear $W$-algebra presentation with a finite number
of generators (the non-linearity in the right hand side is at most quadratic). \par
The construction presented in this work is purely algebraic.  The connection of ${\ell}$-conformal Galilei $(\ell > 1)$ 
algebras to dynamical systems has been, for a long time, an open problem. 
Quite recently, however, in the purely bosonic case, systematic methods (based on non-linear realization and Maureer-Cartan equations) to derive dynamical equations from these algebras, have been developed \cite{GaMa2012,AGM,AGKD} 
(see also \cite{GoKa,AndGo1,AndGo2}). 
So far, the construction of supersymmetric dynamical systems along these lines has not been addressed in the literature. The results presented in this paper pave the way for a systematic investigation, left to future works, of  this class of supersymmetric dynamical systems.

~
\\ {~}~
\par {\Large{\bf Appendix: A comment about the ${\cal N}=4$ supersymmetrization.}}\par
~\par

The steps towards an ${\cal N}>2$ supersymmetrization of the ${\ell}$-conformal Galilei algebra are,
essentially, straightforward. At first one considers the $D$-module reps of the supersymmetry in $0+1$
dimensions (for ${\cal N}=4,8$, see \cite{{pt},{krt}}, they are given by the $(k,{\cal N}, {\cal N}-k)$ field multiplets for $k=0,1, \ldots, {\cal N}$ and with $\lambda, \lambda+\frac{1}{2}, \lambda +1$ as respective scaling dimensions). Next the $D$-module reps of the one-dimensional finite simple superconformal algebras are taken into account; to consider $d$ spatial dimensions  the extra generators $P^{(n)}_i$ and $M_{ij}$, as introduced in Section {\bf 2}, have
to be inserted.\par
Due to new features, the analogy with the ${\cal N}=2$ case stops here. The first new feature is the criticality of $\lambda$. Unlike ${\cal N}=2$, $\lambda$ enters the structure constants of ${\cal N}=4$, allowing to identify the one-dimensional superconformal algebra as $D(2,1;\alpha)$ for $\alpha=(2-k)\lambda$ \cite{kuto}. For ${\cal N}=8$ four one-dimensional simple superconformal algebras ($D(4,1)$ for $k=0,8$, $F(4)$ for $k=1,7$, $A(3,1)$ for $k=2,6$, $D(2,2)$ for $k=3,5$), are identified at the critical values $\lambda=\frac{1}{k-4}$ (see \cite{khto} for details).\par
In a $d$-dimensional space, the scaling dimension $\lambda$ is replaced by the operator ${\widehat{\lambda}}$ given in (\ref{lambdahat}). For ${\cal N}=4$, ${\widehat{\lambda}}$ defines an operatorial-valued $D(2,1;\alpha)$ subalgebra. Since, however, ${\widehat{\lambda}}$ commutes with all generators in
$D(2,1;\alpha)$ it can be treated, for this particular subalgebra, as a $c$-number operator.\par
The other new feature is that the closure of the ${\cal N}=4$ ${\ell}$-conformal Galilei in $d$ dimension
requires, unlike the ${\cal N}=2$ case, the introduction of an infinite number of extra generators. Therefore
the ${\cal N}=4$ superalgebra, even for positive integer or half-integer values of the parameter ${\ell}$, is
an infinitely generated super-Lie algebra. This is implied by the fact that the $R$-symmetry of the one-dimensional superconformal subalgebra is no longer abelian.\par
It should be pointed out, however, that, at least for $k=4$, the ${\cal N}=4$-extension admits a presentation
as a non-linear super $W$-algebra with a finite number of generators, the non-linearity in the right hand
side of the (anti)commutators being at most quadratic. The formulas are rather cumbersome and, being
outside the main line of this paper, will not be reproduced here. They will be presented in a future work.
\\ {~}~
\newpage
\par {\Large{\bf Acknowledgements}}
{}~\par{}~\par 
This work received support from CNPq. The work of N. A. is supported by a grants-in-aid from JSPS (Contract No.23540154).

\end{document}